\documentclass[prb,twocolumn,amsmath,amssymb,floatfix,footinbib,bibnotes,longbibliography]{revtex4-2}
\pdfoutput=1
\usepackage{soul}
\usepackage[colorlinks=true,citecolor=blue,linkcolor=magenta]{hyperref}
\usepackage{multirow}
\usepackage{amsmath}
\usepackage{bbm}
\usepackage{graphicx}
\usepackage{dsfont}
\usepackage{changepage}
\usepackage{fancyhdr}
\usepackage{amsthm, amssymb}
\usepackage{array}
\newcolumntype{P}[1]{>{\centering\arraybackslash}p{#1}}
\usepackage[usenames,dvipsnames]{color}
\usepackage[section]{placeins}
\usepackage[T1]{fontenc}
\usepackage[latin9]{inputenc}
\usepackage[active]{srcltx}
\usepackage{color}
\usepackage{amssymb}
\usepackage{esint}
\usepackage[version=4]{mhchem}
\usepackage[normalem]{ulem}
\usepackage{mathtools}

\newcommand {\apgt} {\ {\raise-.5ex\hbox{$\buildrel>\over\sim$}}\ }
\newcommand {\aplt} {\ {\raise-.5ex\hbox{$\buildrel<\over\sim$}}\ }

\def\titlename{Emergent glassiness in disorder-free Kitaev model: DMRG study on 1D Ladder setting}

\def \authornames{K. B. Yogendra$^{{1}}$, Tanmoy Das$^{{1}}$, G. Baskaran$^{{2,3,4}}$}
\def \affiliations{$^1$Department of Physics, Indian Institute of Science, Bangalore 560012, India \\
\textit{$^2$ The Institute of Mathematical Sciences, CIT Campus, Chennai 600 113, India} \\
\textit{$^3$ Department of Physics, Indian Institute of Technology Madras, Chennai 600036, India}\\
\textit{$^4$ Perimeter Institute for Theoretical Physics, Waterloo, ON N2L 2Y5, Canada} \\}

\begin{document}
\title{\titlename}

\author{\authornames}
\affiliation{\affiliations}


\begin{abstract}
The complete phase diagram of the Kitaev model with a magnetic field remains elusive, as do the experimental results in the candidate material $\alpha$-RuCl$_3$. Here, we study the Kitaev model on a one-dimensional ladder setting within the density-matrix renormalization group (DMRG) method in the presence of a magnetic field at zero temperature. We find five distinct phases with increasing magnetic field, which are characterized by a homogeneous flux phase, the $Z_2$ vortex gas, solid and emergent {\it glass} phase, and finally, a spin-polarized phase. The emergent glassiness is confirmed by calculating correlation functions showing quasi-long range behaviour and ground state fidelity, showing a plethora of energetically accessible orthogonal saddle points corresponding to different flux configurations. This glassy behaviour seems to arise from the slow dynamics of the $Z_2$ fluxes, which is a consequence of the local constraints present in the underlying Hilbert space. This phenomenon can also be explored in other spin-liquid systems where the corresponding low-energy excitations are similarly retarded due to constraints.   
\end{abstract}

\maketitle

\section{Introduction}
Quantum magnetism in crystalline solids and the study of spin liquids is experiencing a resurgence. It is partly due to a remarkable exactly solvable quantum spin model on a honeycomb lattice by Kitaev \cite{KITAEV2006}, followed by an exciting proposal by Jackeli and Khaliullin \cite{Jackeli2009} of experimental realization of Kitaev spin liquid in certain real materials. Several potential Kitaev proximity materials are appearing on the scene \cite{Yogesh2010,Liu2011,Choi2012,Ye2012,Comin2012,Hwan2015,Kubota2015,Williams2016,kitagawa2018}. New experimental results in possible Kitaev systems, such as $\alpha$-RuCl$_3$ \cite{Plumb2014,Sandilands2015,Sears2015,Majumder2015,Johnson2015,Banerjee2016,Do2017,Wagner2022magneto}, continue to surprise us. Beyond basic sciences, developments in quantum spin liquids give hope and pave the way for novel qubits, topological quantum computation, and quantum information science and technology.

The Kitaev model is studied extensively in the presence of the magnetic field in 2D honeycomb lattice \cite{Nandini2019,Pollmann2018,LiangFu2018,MiangLu2018,Trebst2019,Valenti2019}, in ladder setups \cite{Gordon2019,Affleck2020}, and combined with other interactions \cite{Yadav2016,Jiang2019,Kim2020}. There has been a variety of results and proposals, some of which are ubiquitous while others remain active research topics. There is theoretical evidence of $U(1)$ quantum spin liquid (QSL) in the intermediate field regime, with gapless excitations whose nature is still debated (for reviews see \cite{Kee2016,Hermanns2018,Takagi2019Concept,TREBST2022}). Our understanding of the constituent gauge and matter excitations in the Kitaev model with other interactions \cite{Schaffer2012,Craig2012,Oitmaa2015,Janssen2019,Gohlke2018,Batista2021Var,Feng2020,PRADHAN2021} and  external perturbations are gradually evolving  \cite{Berke2020,Batista2022,Aprem2022,Sodemann2022,He2020,Subhro2020}. In particular, the behaviour of the gauge fluxes is not explicitly investigated in the previous numerical studies at finite magnetic fields, and hence their role in the corresponding phases remained unknown. Moreover, given that the dynamics of the flux excitations are restricted by the constraints in the underlying theory, interesting features such as amorphous solid and glass phases can be expected here but remain unexplored in the literature.

The experimental situations similarly remained inconclusive. Experiments have observed half quantization in the thermal Hall effect \cite{Matsuda2018}, and quantum oscillations in in-plane longitudinal thermal conductivity without any observed quantization in the corresponding transverse conductivity \cite{Czajka2021} in $\alpha$-RuCl$_3$ in the intermediate magnetic field region. Another experiment has indicated multiple phase transitions in the same field region based on the anomalies in thermal (both longitudinal and Hall) conductivity \cite{Takagi2022} Evidence of magnetic excitations \cite{Banerjee2018} and phonon anomalies \cite{Dean2021} are also presented in experiments in the same field region (before polarized phase appears). 
More recently, this is roughly the same magnetic field region where an experiment finds a signature of significantly strong and unusual temperature dependence in non-linear susceptibilities ($\chi_2$ and $\chi_3$).\cite{Shivaram2021} Generally, a negative divergence in $\chi_3$ is taken as the signature of spin-glass.  However, in $\alpha$-RuCl$_3$ a {\it  positive} (`$\lambda$'-like) divergence in $\chi_3$, along with other unusual signals in $\chi_{1,2}$, indicates the possible presence of some glass phase and/or excitations with slow dynamics, but not a spin glass.

Emergent glassiness in disorder-free many-body systems is seen, sporadically or otherwise, in many earlier works, although the observed phase was not often associated with glassiness. Intuitively, if the ground state is in proximity to a wealth of local minima due to (say frustration-induced or topological-) degeneracy \cite{Chamon2005}, `emergent disorder' arising from an excessive number of conserved quantities \cite{Moessner2017,Prem2017,Hart2021Logarithmic}, or orthogonal catastrophe near a critical point \cite{Anderson1967,Sachdev1999}, or local constraints or local bath, \cite{Chamon2005,Heyl2021QLM,Feldmeier2019} its dynamics are impeded. In modern calculations, it is also shown that if the Hilbert space is partitioned \cite{Khemani2020,Lee2021} and/or disentangled \cite{Grover2014} into (local) Hilbert space, then the ergodicity is hampered.

The presence of a glass phase in the Kitaev model is highly probable, thanks to its distinctive excitation spectrum. When a local spin flip occurs, it induces the non-local excitations of a pair of $Z_2$ fluxes and/or a flux coupled with a Majorana fermion. However, the mobility of these excitations is limited due to constraints within the underlying Hilbert space. Can these excitations eventually freeze into a glass phase? Furthermore, what role does the flux density play in the potential formation of the glass phase? What is the role of the parent spin-liquid phase in it? In this work, we delve into these questions.

Here, we carry out a DMRG study on the Kitaev model on the 1D ladder at a finite magnetic field and zero temperature. The problem has previously been studied using the DMRG, iDMRG, and exact diagonalization methods, with or without the so-called $\Gamma$ interaction term.\cite{Gordon2019,Sorensen2019,Sorensen2021}. 
While the phase boundaries with the magnetic field strength are reproduced below, the identification of the phases, especially the amorphous and glass phases, remained undisclosed in the previous studies. We find that the phases are described by the flux operator on each plaquette, while interesting features are also observed in the local operators such as spin operator per site\cite{Sorensen2021} and flux operator in half of the plaquette. We find a set of interesting phases with an increasing magnetic field. At low fields, the $Z_2$ gauge flux stabilizes in a spatially homogeneous phase before it tends to crystallize. In the intermediate field region, we spot a robust glass phase determined by random spatial distributions of the $Z_2$ gauge fluxes, with possible gapless excitations. 
The emergence of glass physics is corroborated by the signature results of the correlation functions and quantum Fidelity calculations of the ground state. The dynamics of the glass phase have not been studied in detail because they are computationally expensive to simulate using the DMRG method. The glass phase intervenes in the homogeneous flux phase on one side and a homogeneous polarised phase at a high field. The candidate mechanism for the intrinsic glass phase is that the dynamics of low-energy excitations are restricted by constraints within the underlying Hilbert space, not by the quintessential emergence of conserved charges. As the count of $\pi$-fluxes reaches the half-filling fraction of the plaquette number, the fluxes acquire a near-freezing behaviour. We also conducted the same DMRG analysis on a 4-leg Honeycomb strip with cylindrical boundary conditions. The nature of the phases is the same as that of the 1D ladder except for the quantitative values of phase boundaries. The glassy phase is observed for the range of fields where the U(1) spin liquid was proposed \cite{Nandini2019,Pollmann2018,LiangFu2018,MiangLu2018,Trebst2019,Valenti2019}.

Our remaining article is organized as follows. We present our DMRG method and results in the Kitaev ladder at $T = 0$ as a function of the magnetic field and discuss the emergence of various phases with emphasis on the intrinsic glass phase.

\section{Method}\label{sec:Methods}

\begin{figure}[tb!]
	\centering
	\includegraphics[width=1\linewidth]{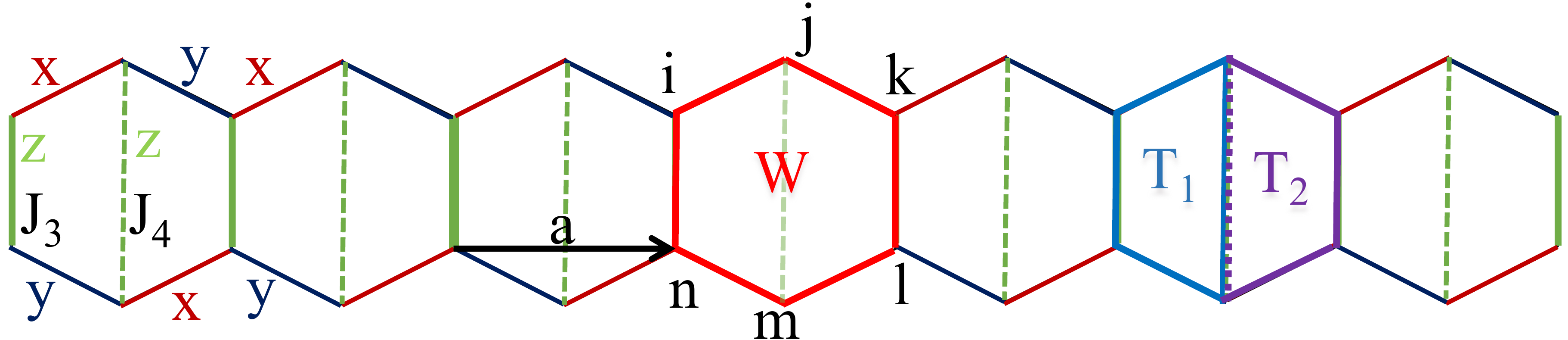}
	\caption{A Kitaev ladder setup that we study here. At each site, we have three nearest neighbor bonds with exchange interactions, $J_{x,y,z}$ between $S^{x,y,z}$, respectively, as in a honeycomb analog. The $J_z$ interactions ($J_{3}$, $J_4$) are kept to be the same as well as different, for comparison. ${\bf a}$ denotes the lattice constant, while $W$, $T_i$ are flux operators defined in the text.}
	\label{fig:Lattice}
\end{figure}

We consider the Kitaev model with the magnetic field (${\bf h}$) along the [111]-direction as
\begin{equation}\label{eq:ham}
	H = \sum_{\langle ij \rangle_\alpha} J_\alpha S_i^\alpha S_j^\alpha - \sum_{i,\alpha}h_{\alpha}S_i^{\alpha}.
\end{equation}
Here $J_{\alpha} > 0$ are bond dependent exchange couplings, $\alpha = x, y, z$. This model is set on the 1D Ladder as shown in Fig. \ref{fig:Lattice}). Each bond has three nearest-neighbor interactions, hence mimicking the setup proposed by Kitaev on a honeycomb lattice. The coupling along the $z$-bond (between the chains) is taken to be staggered, in general, as $J_z=J_3$ or $J_4$ in alternative rungs, see Fig.~\ref{fig:Lattice}.

The spin operator $S_i^{\alpha}$ at each site $i$ can be factorized into matter Majorana fermion ($c_i$) and gauge Majorana fermion ($b_i^{\alpha}$) operators. Then the gauge Majorana operators in the nearest bonds can be combined into a bilinear operator $u_{ij}^{\alpha}=ib_i^{\alpha}b_j^{\alpha}$, which serves as a $Z_2$ gauge field. With this, we can define a flux operator at a six-bond plaquette $p$ as 
\begin{equation}
W_p = S_{i}^yS_{j}^zS_{k}^xS_{l}^yS_{m}^zS_{n}^x =  \prod_{l_p} u_{l_p}^{\alpha},
\label{eq:Wp}
\end{equation}
where $l_p=ij, jk, kl, lm, mn,$ and $ni$ are nearest neighbor bonds. The chosen spin component at a given site is the one present in the outward bond (normal to the plaquette). It turns out that $W_p$ at each plaquette commutes with the Hamiltonian at $h=0$, giving $N$ conserved quantities in both 2D Honeycomb lattice as well as in the 1D ladder. In addition, in the present 1D ladder setting, there are two additional local conserved quantities, which are four-bond plaquette operators as defined by
\begin{eqnarray}
T_{1p} &=& S_{i}^{y}S_{j}^{y}S_{k}^{x}S_{l}^{x} = - \prod_{l_p} u_{l_p}^{\alpha},\nonumber\\
T_{2p}&=& S_{j}^{x}S_{m}^{x}S_{n}^{y}S_{k}^{y} = - \prod_{l_p} u_{l_p }^{\alpha},
\end{eqnarray} 
where $l_p=ij, jk, kl, li$ bonds in the $1p$-plaquette, and so on. These operators are shown in Fig.~\ref{fig:Lattice}. Consequently, $W_p = T_{1p}T_{2p}$ and $\left[ T_{1p}, T_{2p} \right]=0$ \footnotemark[1].  In the ground state, all these conserved quantities assume $W = +1$ and $T_{1p/2p} = +1$, (uniform flux-free phase),\footnotemark[2] giving us an extensive number of conserved quantities. Hence the many-body Hilbert space is made of `trivial' product states of gauge sectors, and matter sectors \cite{Baskaran2007}. This is a $Z_2$ - QSL state \cite{Brenig2017}. The phase diagram of the ladder Kitaev model with $J_{x,y,z}$ has been previously explored in Ref.~\onlinecite{Tao2007}. The lines defined by $J_x-J_y=J_3$, $J_x-J_y=-J_3$ for $J_3=J_4$ exhibit gapless phases characterized by linear dispersions. Furthermore, the $J_4=0$ line also features gapless excitations but with quadratic band touching. The rest of the phase diagram has gapped excitations.
\footnotetext{These four-bond plaquette operators do not commute themselves or with the 2D Kitaev Hamiltonian, but commute with the 1D ladder Hamiltonian in Eq.~1 at $h=0$.}
\footnotetext{Uniform flux free ground state is obtained by fixing the gauge: $u_{ij}^\alpha = +1$ on $\alpha= x, y$ bonds and on the legs, $u_{ij}^z=+1 (-1)$ along $J_3$ ($J_4$)-couplings  \cite{Brenig2017, Satoshi2019, Tao2007}.}

We study Eq.~\ref{eq:ham} at $h\ne 0$ by using the DMRG method for $N=200,300,400$ with cylindrical boundary conditions between the chains and open boundary conditions at the edge. The randomly initialized Matrix product state (MPS) is variationally tuned to the ground state by minimizing the expectation value of the matrix product operator of $H$ in Eq.~\ref{eq:ham} (energy) with bond dimension up to $D \leq 2500$ and truncation error, $\epsilon \sim 10^{-10}$. The DMRG algorithm is implemented using ITensors Library \cite{itensor}. All the results of the main text are with $J_x=J_y=J_3=J_4=1$; see Appendix.~\ref{sec:OtherCouplings} for results of other couplings. The expectation values of any gauge-invariant operators are calculated by contracting the MPO with DMRG-predicted ground state MPS.

We repeat some of the calculations on a four-leg 1D lattice with cylindrical boundary conditions along the armchair direction and open boundary conditions along the zig-zag direction. This geometry is closer to the 2D Honeycomb lattice; see Appendix. \ref{sec:2D}. The salient properties that are presented in the main text for the two-ladder are reproduced in the four-ladder settings.

\section{Results}\label{sec:Results}

\begin{figure}[ht]
	\centering
	\includegraphics[width=1\linewidth]{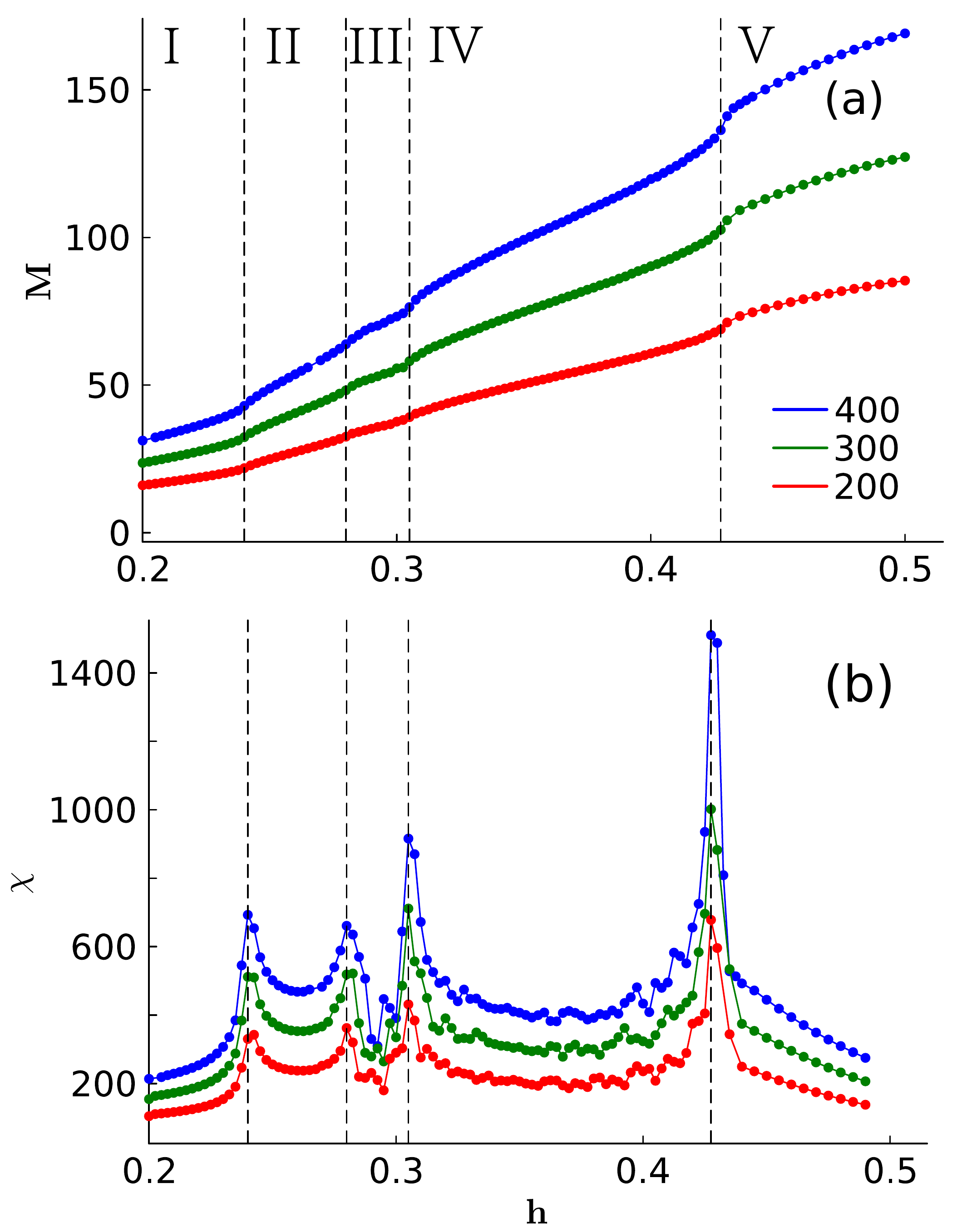}
	\caption{{(a) The spatial average value of the magnetization along the magnetic field direction is plotted as a function of field strength. (b) Corresponding values of the uniform spin susceptibility ($\chi$) are plotted here. Three different colors denote the same calculated values but for three different system sizes $N=400, 300, 200$}. The vertical dashed lines mark the phase boundaries which are located at $h \simeq 0.24, 0.28, 0.3$, and  $0.43$. (The plots are magnified between $h=0.2$ - $0.5$ values for visualization.)} 
	\label{fig:magGapped}
\end{figure}

\begin{figure}[ht]
	\centering
	\includegraphics[width=1.0\linewidth]{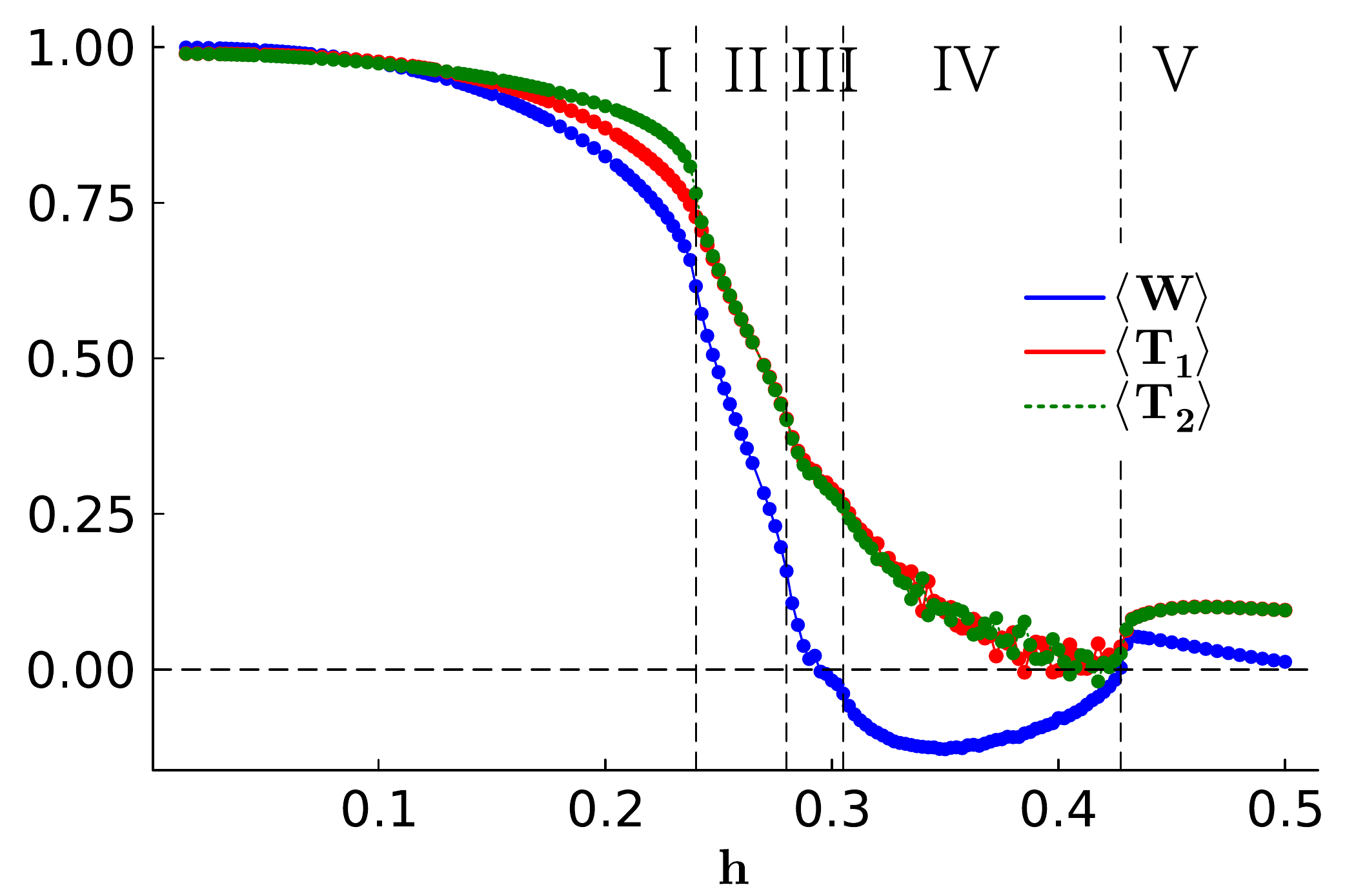}
	\caption{Computed values of the spatial average of the three flux operators $W$, $T_{i}$ are plotted as a function of field strength. The results are shown for a DMRG run on a $400$ site lattice. The vertical dashed lines indicate the same phase boundaries as in Fig.~\ref{fig:magGapped}. The horizontal dashed line marks the $\langle W\rangle =0$ line.}
	\label{fig:AvrgPlq}
\end{figure}

In gauge theories of the present kind, it is often difficult to find the right order parameter (s), especially when there are multiple phases that compete and/or coexist. As $h\rightarrow 0$, we have a non-local multi-linear operator, $W_p$, which acquires a fixed eigenvalue at each site as discussed above. At $h\rightarrow \infty$, the local linear (magnetization) operator $S_{i}^{\hat{h}}$ has a uniform average value in the polarised phase with the easy axis oriented along the field direction $\hat{h}$. There is no obvious way to smoothly interpolate between these two (quasi-) local operators, and a phase transition between them, if exists, evades the Landau theory and occasionally can be classified within the deconfined quantum critical paradigm. Non-local string operators arise as dynamics are introduced in the intermediate magnetic field strength. These string operators bind flux-flux, matter-matter, and/or flux-matter excitations. It is numerically expensive to evaluate their expectation values within DMRG. We will, however, occasionally comment on the possible role of such non-local string operators for the slow dynamics of the glassy phase we obtain here.

\begin{figure}[t] 
\centering
  \includegraphics[width=1\linewidth]{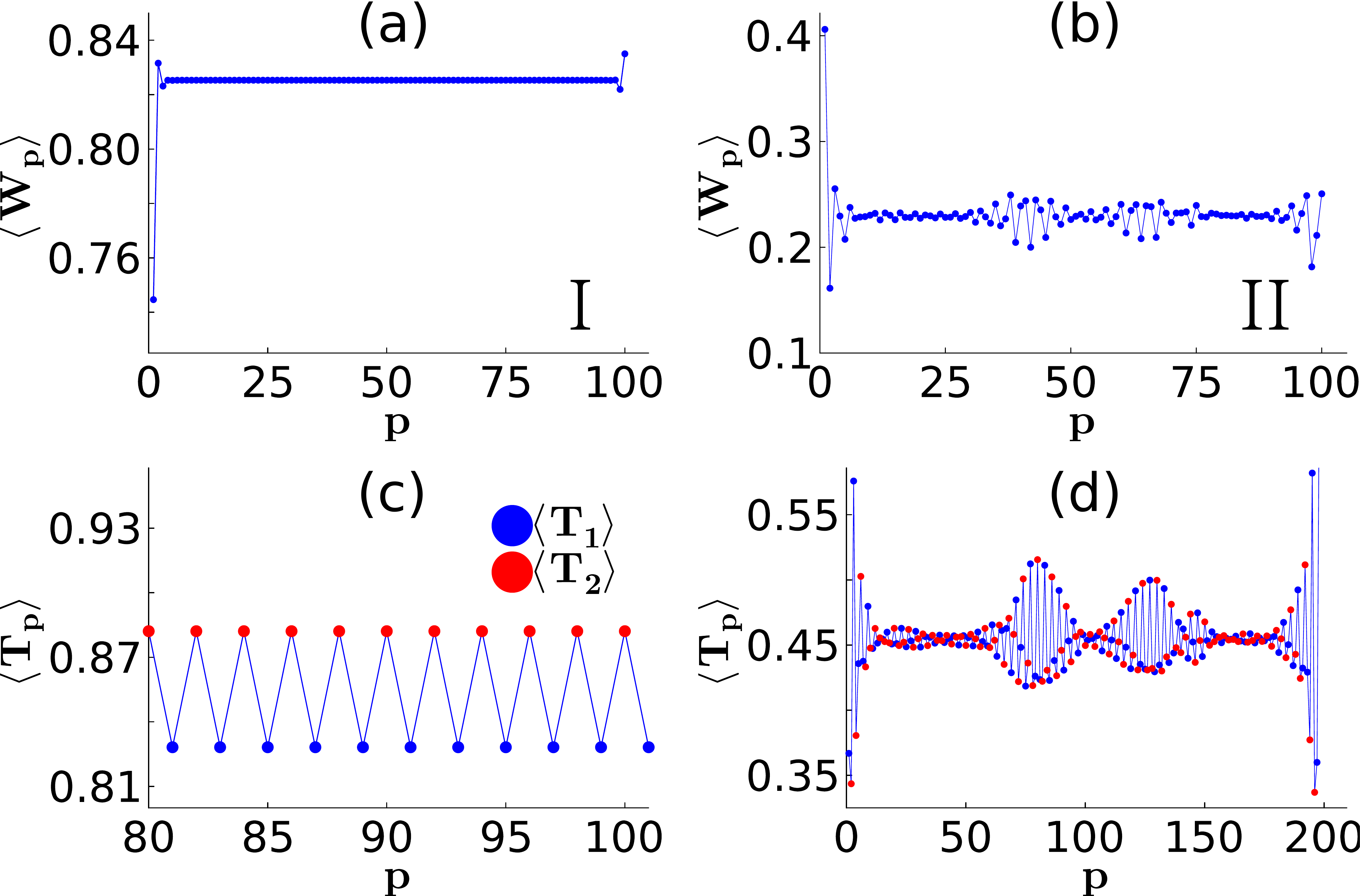}
\caption{The computed value of $\langle W_p \rangle $  are shown for each plaquette $p$ for two different fields (a) $h=0.2$ (b) $h=0.275$, which correspond to Phase I and Phase II. (c),(d)) The values of $\langle T_{\beta p} \rangle$ are shown in the corresponding bottom panel. The $T_{2p} > T_{1p}$ at a $p$ corresponds to $T_{1p}$ flux sitting at the boundaries, and vice versa}
\label{fig:Plq_1}
\end{figure}

We present the spatial average values of the ground-state expectation value $\langle \mathcal{O}\rangle = \frac{1}{N}\sum_l\langle \mathcal{O}_l\rangle$, where $\mathcal{O}_l=S_i^{\hat{h}}$, $l=i$ site index as shown in Fig.~\ref{fig:magGapped}, and $\mathcal{O}_l=W_p$, $T_{1p}$, $T_{1p}$; $l=p$ plaquette index, as shown in Fig.~\ref{fig:AvrgPlq}. In both values of $M=\langle S\rangle$ and $\langle W\rangle$, we observe concurrence of kinks or jumps with increasing magnetic field strength $h$. We denote these finite-field phases by I, II, III, IV, and V. We see in Phase I, a uniform flux value at all plaquettes with the average value decreasing with $h$, and hence we dub it the uniform-flux phase, see Fig.~\ref{fig:Plq_1}. In Phase II, local flux (we will call them $Z_2$ vortex) values begin to deviate from their finite mean value
at a few plaquettes. This is a result of the appearance of the low density of $\pi$ fluxes in the ground state. Phase III appears in the region where the number of vortices is nearly half of the number of lattice sites (half-filling), and $Z_2$ vortices tend to crystallize. Phase IV corresponds to the glass phase with random fluctuations in the $Z_2$ vortices around a zero-mean value. Finally, Phase V corresponds to the uniform polarised phase.        

The magnetization grows near-linearly at all field strengths except in the intermediate region. The uniform spin susceptibility, defined as $\chi= \frac{\partial M}{\partial h}$, shows divergence features at all phase boundaries. The divergence in $\chi$ is most sharp at $h=0.43J$, at the phase boundary between the glass and the polarized phases, possibly indicating a phase transition caused by the long-wavelength collective excitations (magnons). 

\subsection{Uniform and Crystalline phases of fluxes}

The expectation values of flux operators show an intriguing behavior, as shown in Fig.~\ref{fig:AvrgPlq}. Up to $h\approx 0.24J$, we observe a uniform value of $\langle W_p\rangle$, but $\langle T_{ip}\rangle$ obtain staggered mean values between the alternative four-bond plaquettes, as shown in Fig.~4(a) and 4(b), respectively. (The condition for $T_{1p}>T_{2p}$ versus $T_{1p}<T_{2p}$ at a given plaquette depends on the open boundary condition.) Moreover, the uniform value of $\langle W_p\rangle <1$ at all plaquettes suggests that the gauge sector of the ground state can still be approximated to be a product state of local basis, but now the local states have changed from $|+\rangle_p$ at $h=0$ to $\alpha_p|+\rangle_p + \beta_p|-\rangle_p$ for $h>0$, where $W_p|\pm\rangle_p=\pm |\pm\rangle_p$, and $\alpha_p^2-\beta_p^2=\langle W_p\rangle$, $\forall p$. The normalization condition dictates $\alpha_p^2 = (1+\langle W_p\rangle)/2$.  

When the Kitaev model is perturbed, in general, one gets complicated multi-body interactions among Majorana Fermions and Z$_2$ gauge fluxes. Z$_2$ gauge fluxes become dynamic and acquire finite effective masses\cite{Aprem2022}. Further, open string operators carrying Majorana fermion modes (both $b_i^{x, y, z}$ and $c_i$) at their ends also have expectation values in the ground state. Study of open strings using DMRG at finite fields is cumbersome. Elaborated discussion on these string objects at finite fields and their role in dynamics is presented in Appendix.~\ref{sec:Strings}. There are excitations due to $T_{ip}$ fluxes whose energy scale is $<10^{-3}J$. But in the uniform $\langle W_p\rangle$ phase, we find that $\langle T_{1p}\rangle > \langle T_{2p}\rangle$ for $T_{1p}$ sitting at the boundaries ($\langle T_{1p}\rangle < \langle T_{2p}\rangle$ for $T_{2p}$ at boundaries). In addition, both the $W$ and $T$ flux pairs are also virtual excitations. These excitations induce further neighbour hopping to the matter fermions. The density of these excitations increases as a function of $h$, changing the magnitude of the $\langle W \rangle$. The energy gap to create flux-pair is $\leq 10^{-3}$. Similarly for $T-$flux pair, it is $\leq 10^{-5}$ (see, Appendix.~\ref{sec:Gaps}). There are long-wavelength collective excitations, in which $\alpha_p$ (i.e. $\langle W_p\rangle$) varies slowly across the lattice but with a gap which scales with the system size. Finally, single matter Majorana excitations appear at higher energy.  

A single $Z_2$ vortex creation in the uniform flux case at a six-bond plaquette, i.e. changing $W_p$ from $+1$ to $-1$ costs energy $E\sim 0.24J$. Therefore, for $h > 0.24J$, $W_p$ vortex creation is energetically feasible. In the dilute limit, the vortices start to proliferate in the lattice like a vortex gas or liquid phase, which is Phase II in our phase diagram. 

\begin{figure}[t] 
\centering
  \includegraphics[width=1\linewidth]{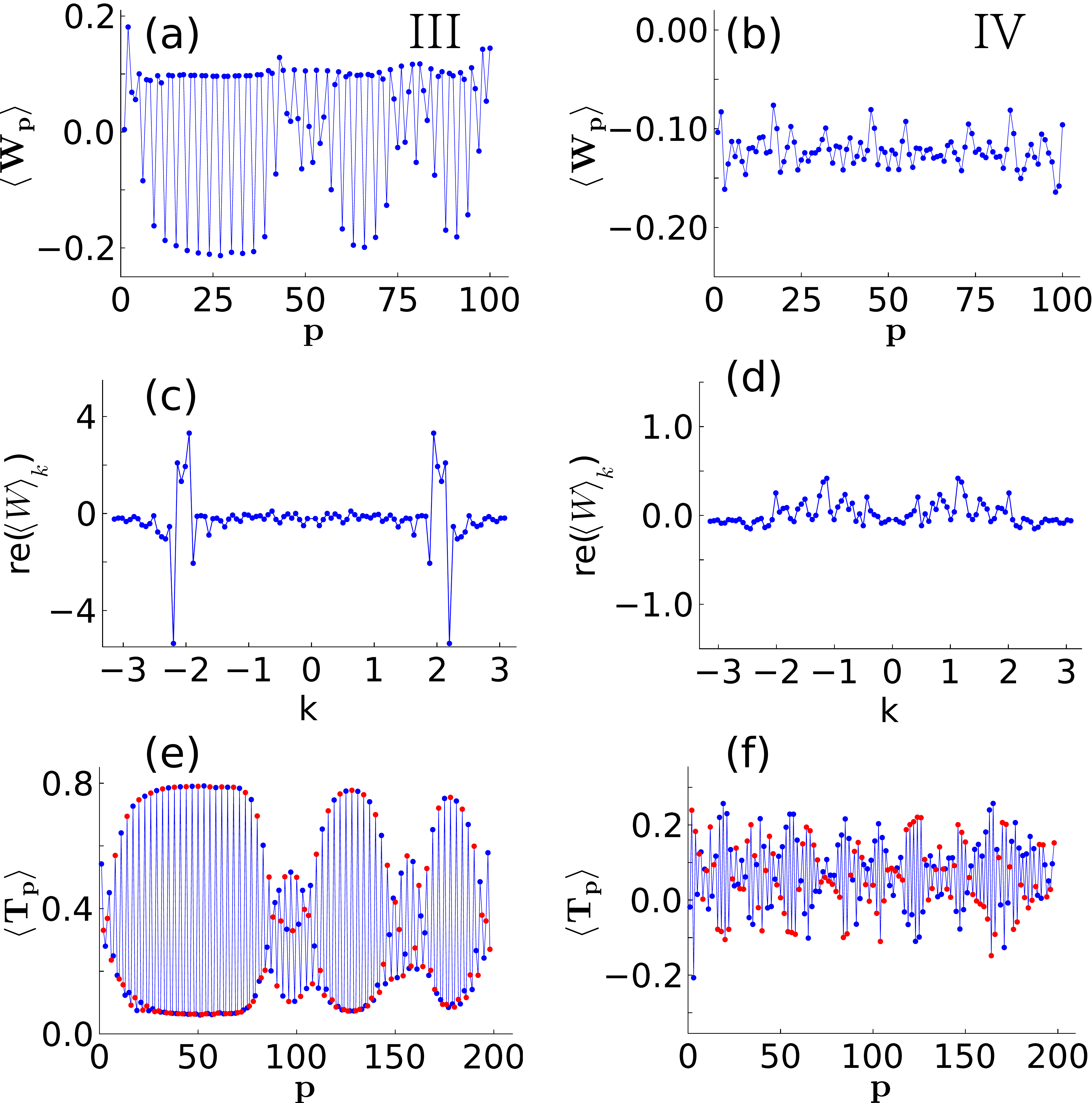}
\caption{Similar to Fig.~\ref{fig:Plq_1}, but here the results are shown at two representative fields of Phase III ($h=0.2975$) and Phase IV ($h = 0.365$). In the middle panel ((c),(d)), we plot the real part of the Fourier transformation of $\langle W_p \rangle$ with wave vector $k$.}
\label{fig:Plq_2}
\end{figure}

With further increase of the field strength, by $h \geq 0.28J$ there is a tendency for the vortices to crystallize, as shown in Fig.~\ref{fig:Plq_2}(a). This is Phase III. Here $W_p=\pm1$ plaquettes are nearly equal in number, giving $\langle W\rangle\rightarrow 0$, which is close to half-filling. In this case, the vortices are `frozen' to the lattice site with alternating plaquettes having opposite $W_p$ sites, see Fig.~\ref{fig:Plq_2}(a). This phase is analogous to a density wave order in a correlated fermionic insulator or hard-core bosonic insulator at half-filling. The vortex lattice formation is evident in the dominant value of the Fourier component of the flux operators at a single wavevector as shown in Fig.~\ref{fig:Plq_2}(c). Slightly away from the half-filling on both sides, we observe here a few wavevectors and quasi-long-range correlation functions. which suggests an amorphous behavior.

\begin{figure}[ht] 
\centering
  \includegraphics[width=1\linewidth]{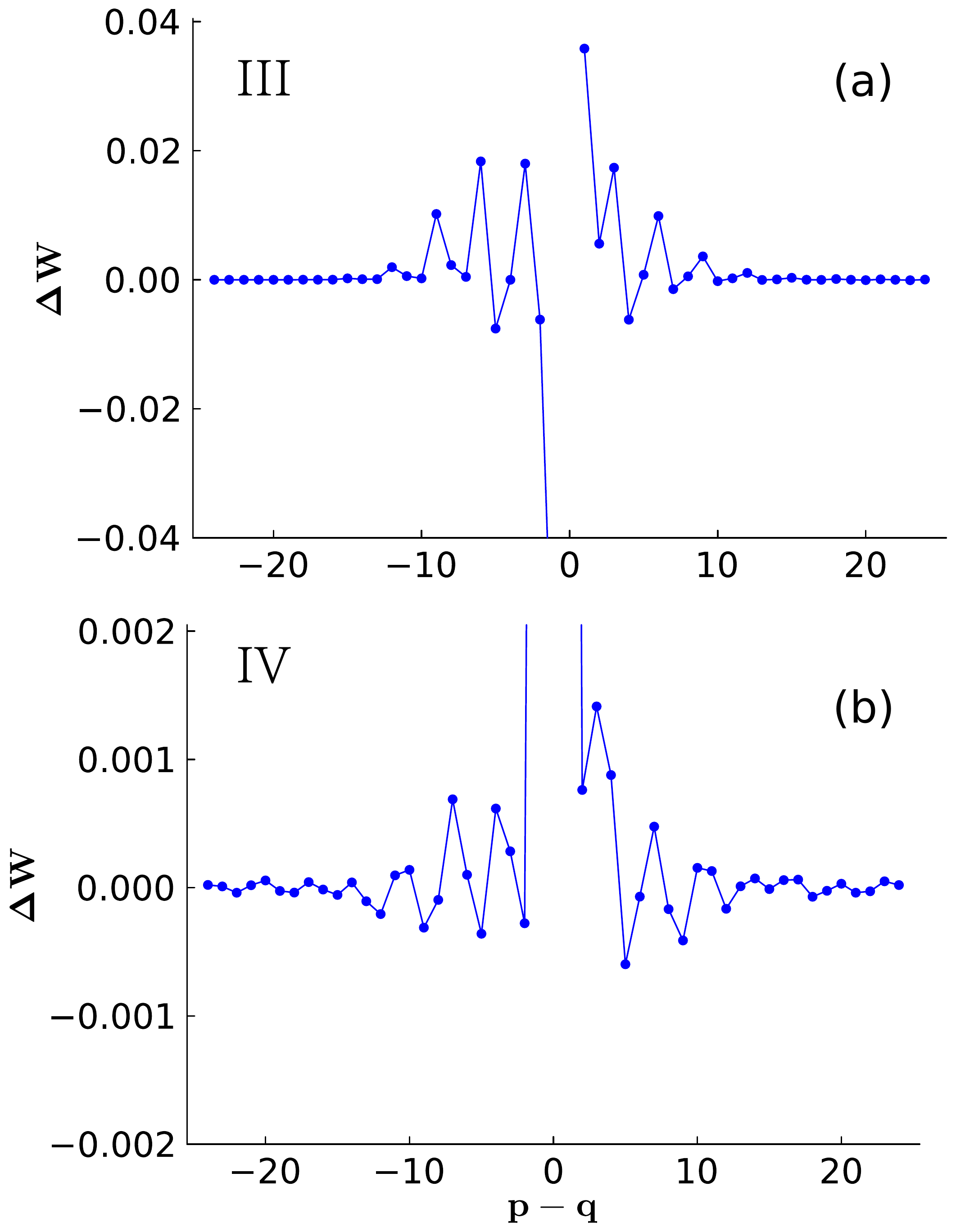}
\caption{We plot the correlation function of the flux operator $\Delta W_p$ with $p=50$ for (a) $h=0.295$, and (b) $h = 0.365$.}
\label{fig:Corr}
\end{figure}

\subsection{Emergent Glassiness}
An amorphous crystal is a precursor to glassiness and may be at play in the present case as well. The energy to create a single $W_p$ flux in the crystalline phase is  $\sim 0.05J$ (assuming uniform crystal for this estimation, see Appendix~\ref{sec:Gaps} for more details). Therefore, at $h>0.3J$, we enter into the dense vortex region (Phase IV). The large value of the $Z_2$ vortex density is evident in the $\langle W\rangle \leq 0$ value shown in Fig.~\ref{fig:AvrgPlq}. Because of this high density, any small local fluctuation tends to impede the ordering of the entire lattice, and hence a {\it glassiness} arises. 

We calculate the correlations of $W_p$, quantifying the fluctuations from its mean, as $\Delta W =\langle W_pW_{q}\rangle-\langle W_p \rangle \langle W_{q}\rangle$, where the expectation value is calculated with respect to the MPS ground state. The value of the correlation of the fluxes is $\sim 10\%$, and the correlation length extends up to 20-30 plaquette distance on both sides as shown in Fig. \ref{fig:Corr}(d). Furthermore, this quasi-long correlation length in the glass phase is larger than its precursor crystalline phase. This is in contrast to a solid-to-liquid phase transition where the correlation length decreases in a liquid phase. This is one aspect of the glassiness that distinguishes phase IV from it being the liquid phase. 

In this phase, the local spin operators exhibit an incommensurate order, as also reported earlier.\cite{Sorensen2021} Different incommensurate orders in finite lattice give closely lying energy states, 
but fail to become orthogonal to each other. On the other hand, as we will see below via the Fidelity calculation, different local minima states are completely orthogonal to each other, suggesting that the glass state is not governed by the incommensurate local spin but by the flux operators.

Furthermore, we have also checked that the phase has a non-zero central charge, signaling gapless excitations. Note that this is the approximate range of fields where gapless $U(1)$ QSL state is proposed in the 2D Honeycomb Kitaev model \cite{Nandini2019,Pollmann2018,LiangFu2018,MiangLu2018,Trebst2019,Valenti2019}. The convergence of DMRG minimization in this range of fields is slow compared to time scales for other phases (see the Fidelity result for the corresponding interpretation).

Referring to the definition of $W_p$ in Eq. \ref{eq:Wp}, it is easy to associate the fluctuation of $\langle W_p\rangle$ with the quantum fluctuation of the spins. This sets the present glass physics apart from the classical glassy phase of frozen spin configurations. Note that apart from single flux productions, there are also non-local flux pairs that are connected by Wilson operator $(W_p)^n$ - which in the spin operator form takes a string operator. This automatically generates $n-$point spin-spin correlations in this system. Definite a $n^{\rm th}$-order uniform susceptibility $\chi_n\sim\partial^n M/\partial h^n$, we have checked that the second and third-order susceptibilities in this region are large and more chaotic as a function of the magnetic field. Note that in a Gaussian fluctuation theory, the third and higher-order susceptibilities vanish, as we also find in the other phases. But in Phase IV, we find significant enhancement of the mean square values of the second and third-order susceptibilities in the range of $\mathcal{O}\left(10^2\right)$ to $\mathcal{O}\left(10^3\right)$.

For the high magnetic fields, Phase V is trivially polarised along the $[111]$ direction. The fluxes are half of the plaquettes with $\pi-$ fluxes resulting in $\langle W_p \rangle = 0$ and $\langle T_{1p/2p} \rangle = 0$ in every plaquette uniformly. 

There is no glassy phase observed for the ferromagnetic couplings, i.e. $J_x=J_y=J_3=J_4=-1$. Here we find two phases: the uniform flux phase at a small magnetic field region and the polarised phase with uniform $\langle W_p \rangle=0$ (and $\langle T_{1p/2p}\rangle=0)$.

\subsection{Robustness of results with other Lattice Settings}
We repeat the DMRG calculation in a 2D lattice strip via the four-leg Honeycomb lattice with cylindrical boundary conditions; see Appendix.~\ref{sec:2D} for more details. We find four phases (Fig. \ref{fig:Avrg2D}), where phase II and phase III are not distinguishable within the finite system size calculation. More importantly, the glass phase is reproduced here.

We also repeat the DMRG calculation in 1D ladder for $J_4=0$ with other parameters fixed at 1. This creates open boundary conditions between the chains. The result is presented in Appendix~\ref{sec:OtherCouplings}. We find two phases: at $h>0$ we immediately find a crystalline phase (phase III) and the uniform polarized phase (phase V). The glass phase is absent here.

\section{Phase transitions and Fidelity}

\begin{figure}[htb!]
	\centering
	\includegraphics[width=1\linewidth]{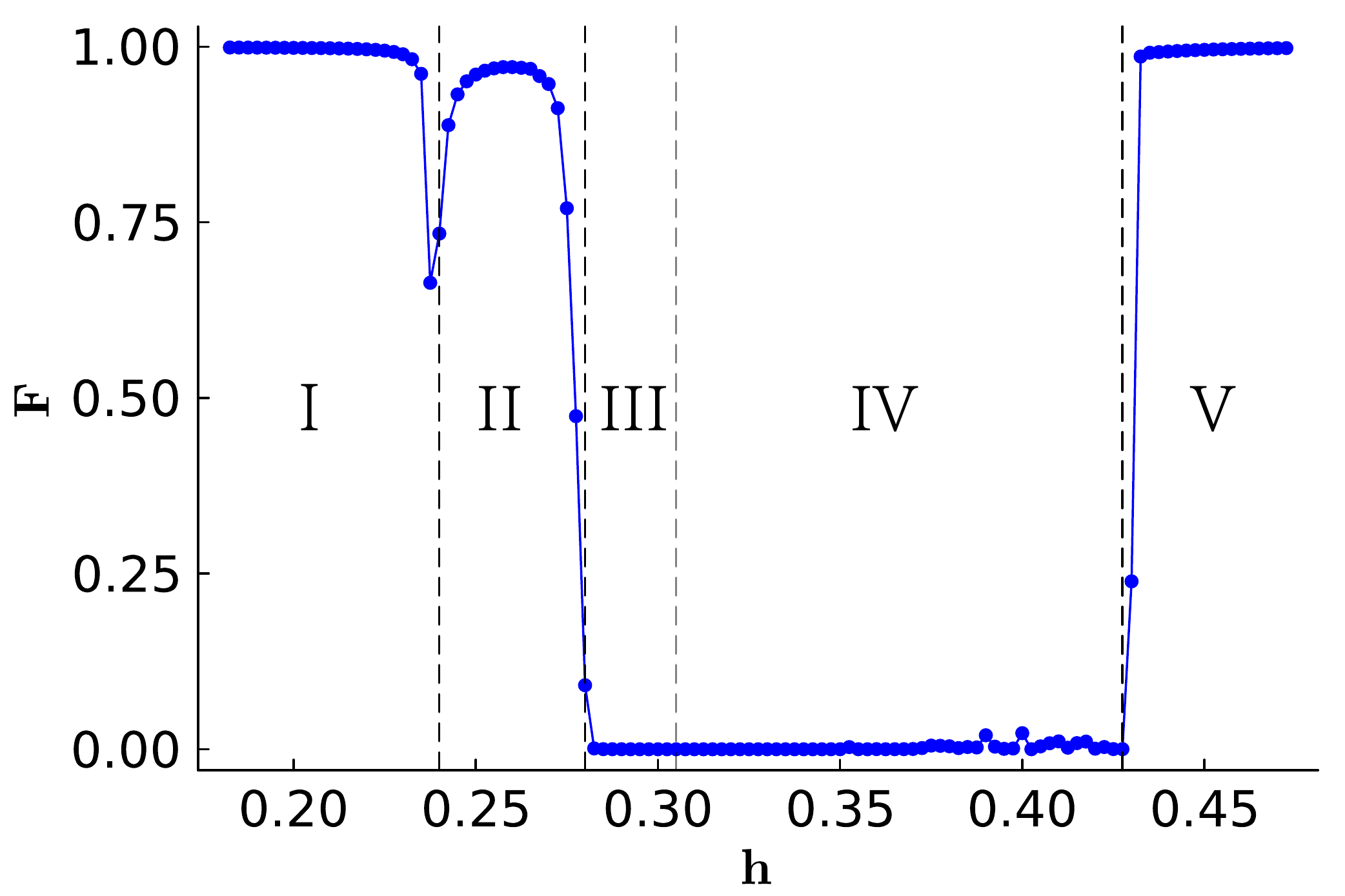}
	\caption{We plot the quantum Fidelity (defined in the text) with magnetic field for $N=400$ lattice sites. The vertical dashed lines indicate all five phase transition points, which coincides with Figs. \ref{fig:magGapped} and \ref{fig:AvrgPlq}.}
	\label{fig:Fidelity}
\end{figure}

In the absence of a well-defined local order parameter, characterizing phase boundaries and phase transitions becomes challenging. In such a scenario, we can study how different variational ground states are mutually orthogonal as a function of the control parameter. This information can be obtained by the quantum fidelity analysis. 

The quantum Fidelity is defined as $F(h)= \left|\langle \psi_0\left(h\right)| \psi_0\left(h+\delta h\right) \rangle \right|$, where $|\psi_0 \left(h\right) \rangle$ is the ground state vector obtained from the DMRG calculation at $h$ \cite{Zanardi2006,Vidal2008,Zhou2008,Yang2008,Mukherjee2012}. It is now evident that if the states $|\psi(h)\rangle$ and $|\psi(h+\delta h)\rangle$ are linearly dependent, we have $F\rightarrow 1$, and if they are completely orthogonal we get $F\rightarrow 0$, and any value between them measures the overlap between the two wavefunctions.\cite{Castelnovo2010} The fidelity $F$ vanishes in two major scenarios: due to gapless excitations and emergent glassiness. As the system is tuned to a new configuration, and if the corresponding state is orthogonal to the preceding one, the fidelity between the two states vanishes. This phenomenon is known as the orthogonality catastrophe, as proposed by Anderson in free Fermion systems. It is an infrared catastrophe that arises from gapless excitations. However, fidelity can also vanish due to emerging glassiness. In this case, the system becomes trapped in some metastable local minima whose states are mutually orthogonal to each other. We discuss below why the glass phase is the origin of the $F=0$ value in our case.


As shown in Fig. \ref{fig:Fidelity}, we see that $F\rightarrow 1$ in both the uniform phases of flux (Phase I) and of spin (Phase V), suggesting a unique ground state in these phases. $F$ sharply decreases at the phase boundary between Phase I and II, implying that the vortex gas phase is separated from the uniform phase by a phase transition. Within the Phase II region, the Fidelity does not completely reach 1, suggesting the presence of configurations that partially overlap with the chosen ground state. 

The most exciting feature is obtained in Phase III (amorphous vortex crystal) and Phase IV (vortex glass) where $F=0$. This clearly indicates the presence of a plethora of local minima whose wavefunctions are orthogonal to the chosen ground states. These local minima are not degenerate, as in the case of an infrared catastrophe, but lie within the energy fluctuation scale provided by the magnetic field. Additionally, we repeated the DMRG runs with different random initial configurations at a fixed value of the magnetic field. In each iteration, the obtained ground states are orthogonal to each other. We have performed the ED calculations with system sizes 6, 10, and 14. And found that the results qualitatively agree with the DMRG; see Appendix~\ref{sec:ED} for the details. These findings strongly suggest that the orthogonality is unlikely to be driven by gapless excitations and rather points towards the stabilization of a glass phase.

The fidelity is consistently zero in both Phase III and Phase IV, which aligns with expectations since an amorphous solid serves as a precursor to a glass phase. In an amorphous solid, a domain of fluxes has a metastable ground state, while in the glass phase, the domain size reduces to a single flux. Orthogonal ground states obtained from different DMRG runs for phase-III have different domain wall structures. Since the flux configurations vary for different initial conditions in DMRG runs, the size of domains varies from one another. However, the size of the largest domain scales proportionately with the system size.

\section{Discussion}

A key feature of the QSLs, in general, and the Kitaev model, in particular, is the presence of constraints on the Hilbert space for low-energy excitations. The emergent matter excitations typically enjoy an enlarged Hilbert space than the physical Hilbert space of the spin operator provided by the Hamiltonian. This restriction on the available Hilbert space limits the dynamics of the excitations on the phase space. 

In this DMRG study, we observed the glassy phase at the intermediate field range on the 1D ladder and then in a 4-leg Honeycomb strip with a cylindrical boundary. Previously, various numerical studies have suspected intrinsic glassiness or slow dynamics in general as a function of temperature and applied field. Localization features are observed for a range of different couplings in the Kitaev ladder model at low field \cite{Brenig2021}, indicating that the localization behavior is beyond uniform flux approximations at low fields. A non-ergodic phase is also proposed in the 2D Kitaev model under quench with skew magnetic field \cite{Heyl2021} and without the field for anisotropic couplings \cite{Rademaker2019}. Recall that the exact solution of the Kitaev model at the zero field is a result of $N$ local conserved quantities (flux operators) in a honeycomb lattice of $N$ plaquettes. As claimed in Refs.~\onlinecite{Khemani2020,Lee2021}, the local conservation constraint leads to the shattering of the full Hilbert space of dimension 2$^{2N}$ into 2$^N$ sectors of equal dimension. Each sector defines a 2$^N$ dimensional Hilbert space of free neutral fermions. This perfect partitioning of Hilbert space and consequent superselection impairs ergodicity and favors many-body localization. 

Any external perturbation that directly couples to the local spins generates excitations of flux pairs or flux-Majorana bonding. These topologically protected excitations have restricted dynamics in the lattices and are impervious to annihilation by temperature or local defects \cite {Hart2021}. However, the emergence of a glass phase, specifically a flux glass rather than a spin glass, necessitates further fine-tuning. In our study, we observe that as the number of fluxes reaches half-filling, the disorder configurations of these fluxes form local energy minima with corresponding orthogonal quantum states. These local minima possess slightly different energies, comparable to the magnetic field's strength, and do not give rise to a degenerate manifold. Consequently, an infrared catastrophe does not occur; instead, the system freezes into one of these local minima configurations. Since the flux operators are products of spin operators within a plaquette, the flux correlation function corresponds to a many-spin correlation function. Measuring such a correlation function is currently infeasible using existing experimental techniques. However, indirect measurements can be achieved through experiments involving field quenching and non-equilibrium analyses. For instance, femtosecond laser pulses can be employed to probe the glassiness present in Kitaev spin liquid materials.

Extending these ideas to the RVB state, a few conjectures can be made for future studies. A local spin flip at a site produces two spinons, which separate away during time evolution. However, spinons as sources of emergent gauge fields, carry gauge fluxes \cite{baskaran1988gauge}; sometimes both electric and magnetic charges, called Dyons \cite{Affleck1986,Baskaran2003Skyrmion}. 
Net gauge fluxes created by the spin operators are zero, even though the spin operators are gauge invariant. Flux attachment endows spinons with fractional exchange statistics in 2 dimensions \cite{Read1989}. This is also transparent in Kalmeyer-Laughlin chiral spin liquid state \cite{Kalmeyer1987,Laughlin1990} and later works, where low energy spinon carries a Vison or Meron (half-Skyrmion) \cite{Baskaran2003Skyrmion} or SU(2) gauge fluxes \cite{Lee2000}. The restrictions on the spinon dynamics lower its kinetic energy. Whether these spinons can freeze to form a glass phase in a U(1) spin liquid phase remains to be investigated in future studies.

\section{Conclusions}
Our detailed DMRG study on the 1D Kitaev model with a magnetic field reveals an intriguing phase diagram with five phases, and among them, we discover a glass phase. All these five phases are also obtained in the Kitaev-$\Gamma$ model in previous studies\cite{Gordon2019,Sorensen2019, Sorensen2021}. We have found an intriguing intrinsic glassiness in a part of their phase diagram when $\Gamma = 0$. It is likely that glassiness is present in other regions of the phase diagram which remains to be investigated in the future. Furthermore, we are able to segregate between the vortex gas, crystal, and glass phases in the otherwise known $U(1)$ QSL phase, due to the detailed analysis of the vortex operators as well as the Fidelity calculation. We find evidence of gapless excitations in the vortex glass phase but not in the gas and crystal phases.

How robust is our phase diagram beyond a two-leg ladder geometry and beyond the limitations of the DMRG studies? A complete answer to this question is not known in the community. We have, however, repeated the DMRG calculation on a four-leg ladder geometry as given in the Appendix. Here, we find four Phases: Phase I, Phase III, Phase IV, and Phase V. This means the boundary between Phase II (vortex gas) and Phase III (vortex crystal) is not discernible. However, the vortex glass of present interest is well reproduced.   

There are now numerical software available for finite temperature calculation within DMRG and Tensor network formalism. Future extension of our calculation to finite temperature will shed light on the possibility of a BKT-like physics for $Z_2$ vortex as well as the stability of the glass phase to thermal broadening. 

\textit{Note:} As we were finishing this manuscript, we came across an interesting paper from Zheng Yan et al. \cite{sachdev2023}, where they report numerical findings of emergent glassiness in disorder-free Rydberg atom arrays in 2D. 

\section*{Acknowledgements}
GB thanks B. Shivaram for the discussion on experimental results from his group on anomalous non-linear susceptibility in $\alpha$-RuCl$_3$ Tarun Grover and Mathew Fisher for discussions. We thank Vijay Shenoy for suggesting the Fidelity calculation. GB acknowledges continuing support from the Institute of Mathematical Sciences, the Indian Institute of Technology in Chennai, India, and the Perimeter Institute for Theoretical Physics at Waterloo, ON, Canada. GB's research at Perimeter Institute is supported by the Government of Canada through Industry Canada and by the Province of Ontario through the Ministry of Research and Innovation. TD acknowledges research funding from S.E.R.B. Department of Science and Technology, India, under I.R.H.P.A Grant No.
IPA/2020/000034 and acknowledges the computational facility at S.E.R.C. Param Pravega under NSM grant No. DST/NSM/R\&D HPC Applications/2021/39. KBY thank ICTS for the accommodation during the program "Frustrated Metals and Insulators" (code: ICTS$/$frumi2022$/$9).

\appendix


\section{Phases for Different Couplings}\label{sec:OtherCouplings}
\begin{figure}[htb!] 
\centering
  \includegraphics[width=1.0\linewidth]{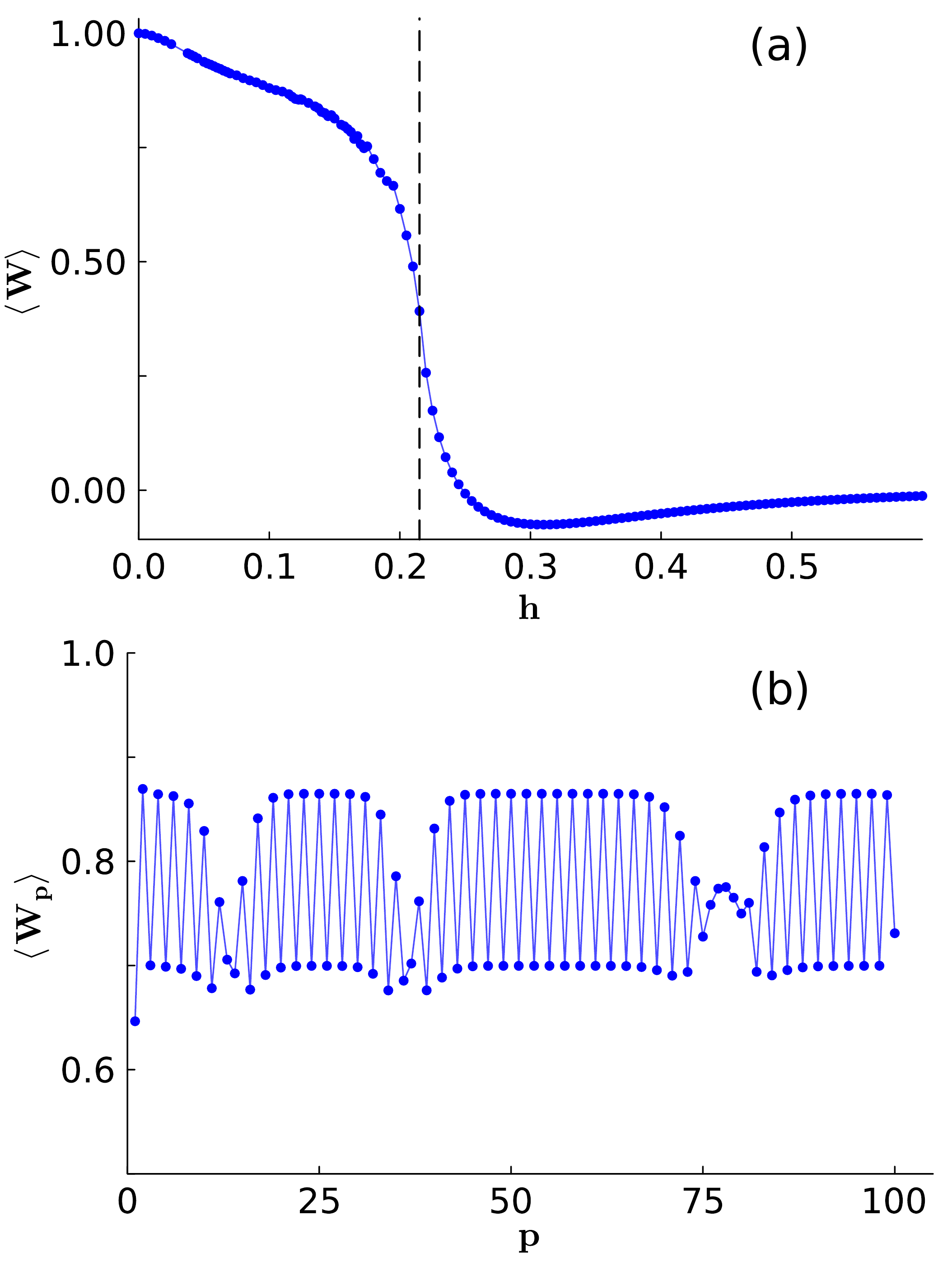}
\caption{(a) Average of $\langle W_p \rangle $ for $N = 400$ as a function of magnetic field for the coupling $J_4=0$ with $J_x = J_y= J_3 =1$. The phase boundary is indicated with a vertical dashed line. (b)$\langle W_p \rangle $ as function of plaquette number, $p$ at field $h=0.165$.}
\label{fig:BWPhase}
\end{figure}
In the main text, we presented results for $J_x=J_y=J_3=J_4=1$.  $J_4=1$ imposed a cylindrical boundary condition perpendicular to the ladder, leaving no bond indices to be open. Now we set 
$J_4=0$, which gives alternating sites to have open bonds in the matrix product state. The energy dispersion of matter fermions in the ground state with $h=0$ is gapless and quadratic.  For $h>0$, only two phases are present with a phase boundary at $h \approx 0.215$, which is distinguished by a cusp in the magnetization plot (not shown). The average value of the $W_p$ operator as a function of the magnetic field is shown in Fig. \ref{fig:BWPhase}(a). The low field phase is a crystalline phase of $\langle W_p\rangle$,  as shown in Fig. \ref{fig:BWPhase}(b). This phase is same as Phase-III for $J_4=1$ given in main text. The high-field phase is a polarized phase with $\langle W_p \rangle = 0$ at all the plaquettes. There is no glass phase observed here.

Two phases with phase boundary around $h \approx 0.25$ as a function of magnetic field are found with couplings $J_x=2, J_y=J_3=J_4=1$, where the ground state dispersion of matter fermions at $h=0$ is gap-less and linear. No structural difference is observed in flux configurations in both phases with $\langle W\rangle$ decreasing smoothly with increasing $h$. 

\section{Towards 2D: Results on 4 Leg model}\label{sec:2D}
\begin{figure}[ht] 
\centering
  \includegraphics[width=1.0\linewidth]{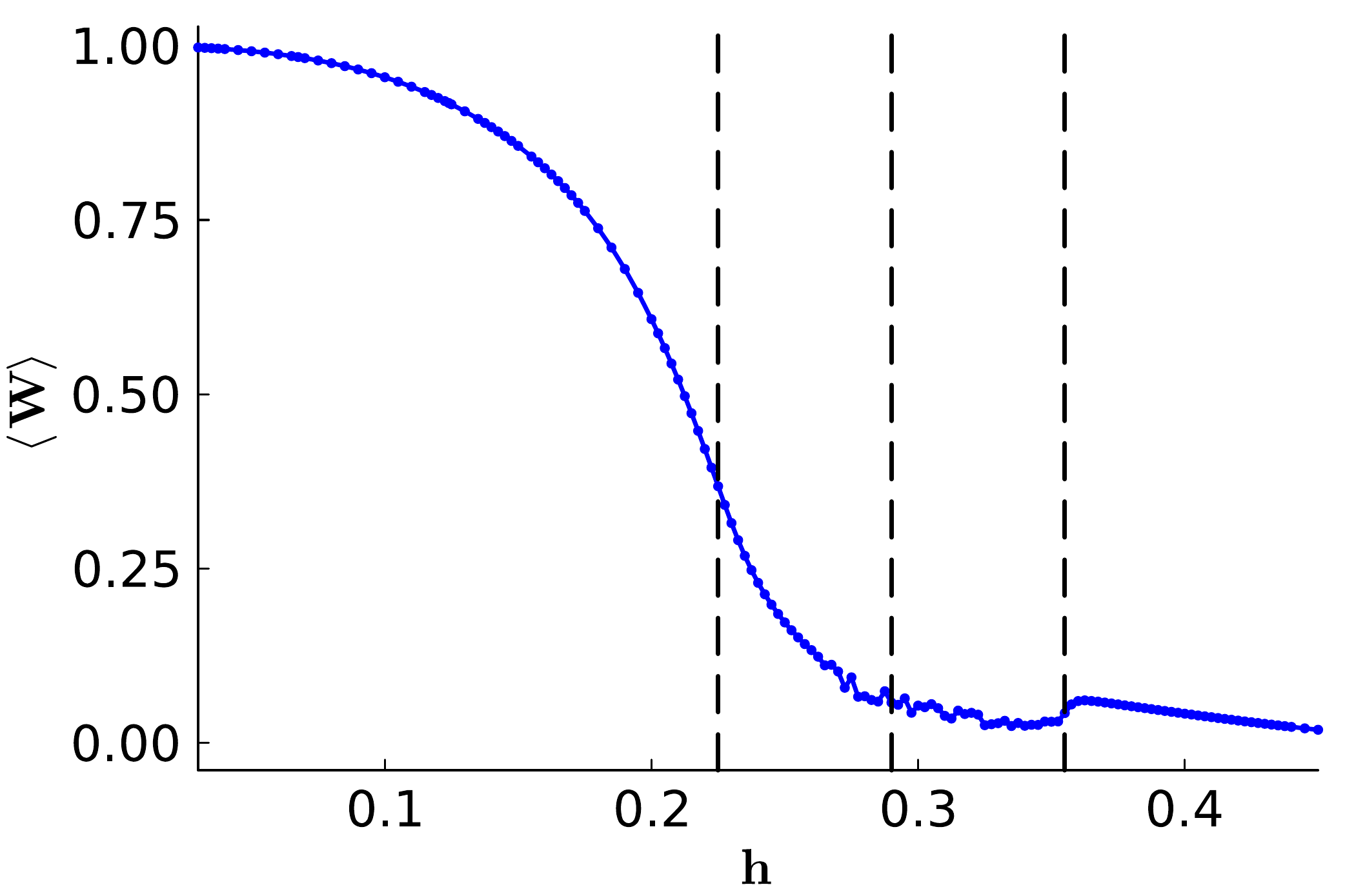}
\caption{The spatial average values of $\langle W_p \rangle $ as a function of the magnetic field for the four-leg Honeycomb lattice with system size $N=52$ sites . The vertical dashed lines pointing the fields of phase transitions identified from the magnetic susceptibility.}
\label{fig:Avrg2D}
\end{figure}

To find out the robustness of the phase diagram presented in the main text on a two-leg DMRG calculation, we repeat the calculation on a four-leg ladder with cylindrical boundary conditions of system size, $N=52$ sites and with bond dimension $D \le 1000$, truncation error, $\epsilon \approx 10^{-10}$. We reproduce four phases as presented in Fig. \ref{fig:Avrg2D}. Those four Phases are Phase I, Phase III, Phase IV, and Phase V of the 1D ladder results presented in the main text.

The different phase boundaries are identified from the magnetization values with smaller steps of the magnetic field than seen in previous studies, see for example, Ref. \cite{Nandini2019}. The average expectation of the plaquette operators as a function of the magnetic field is shown in Fig. \ref{fig:Avrg2D} agreeing with the previous findings \cite{Pollmann2018}.

\begin{figure}[ht] 
\centering
\includegraphics[width=1.0\linewidth]{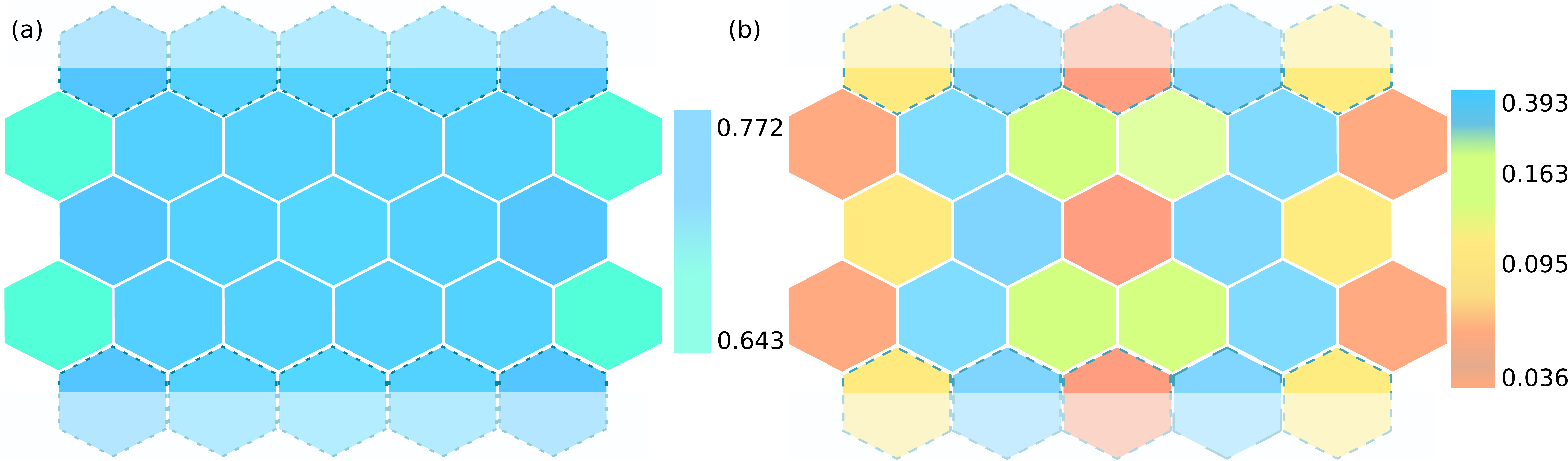}
\caption{$\langle W_p \rangle $ at (a) $h=0.18$ and (b) $h=0.26$, the emergence of periodic ordering in flux configurations is observed at this field. The colour bar indicates the strength of the flux in a given plaquette. The shaded plaquettes are connected for the periodic boundary in the cylinder geometry.}
\label{fig:2Dfluxes}
\end{figure}

Even though the system size is small and prone to boundary effects along the legs, we reproduce four phases as seen in the 1D ladder. Here the boundary between Phase- II to Phase- III is not explicit within the finite size calculations. The low field phase, Phase-I, for fields up to $0.23J$, has uniform flux configurations, as shown in Fig. \ref{fig:2Dfluxes}(a). Then, in Phase III, for a range of fields $0.23 < h < 0.29$, the proliferation of the dynamically generated fluxes into the ordered configurations is observed, see Fig. \ref{fig:2Dfluxes}(b). For fields above $0.29$, the randomly distributed fluxes without any order are observed up to $0.36$ in the proposed U(1) spin liquid region. The flux configurations in the polarised phase are with $\langle W_p \rangle =0$ in all the plaquettes uniformly. 

\section{Estimation of Gaps}\label{sec:Gaps}
The gap to the excited state with $\pi-$ fluxes from uniform flux phase at $h=0$ is estimated by Exact Diagonalization. The ED calculations are done with matter Majorana fermions by fixing the gauge in accordance with uniform flux configuration. The $Z_2$ vortex gap for creating a single $W$ or $T$ is, in principle, calculated by keeping two $W$ or $T$ $\pi-$ fluxes infinitely far apart. In finite-size calculations, that is approximately the y-axis intercept of the plot: gap versus $1/d$, where d is the distances between the two $\pi-$ fluxes with systems sizes, $\left(2 d\right)$. The $\pi-$ flux pair can be created by changing the bond operator from  $u^z_{\langle ij \rangle} = +1$ to $ -1$ on $z-$ bond common to the two adjacent plaquettes. Further creating a series of adjacent $\pi-$fluxes either to the right or left of already created flux-pair for separating those initially created $\pi-$fluxes accordingly. The gap to single $W$ vortex is $\approx 0.24$ and for $W$ flux-pair, $\leq 10^{-3}$. In case of $T$ plaquettes, it is $ \leq 10^{-3}$ for single vortex and $\leq 10^{-5}$ for flux-pair.

The ordered superlattice flux configuration at the finite field strength is approximated to uniform crystal for estimation of the gap. Further, it is approximated as follows: the high $\langle W_p \rangle$ value in the plaquette p to $+1$, the lower one to $-1$. With this approximated flux configuration, the estimated gap to the single vortex is calculated following the same approach as for the uniform flux case. The gap to a single $W$ vortex is $\approx 0.05$ in this approximated uniform crystallised flux configuration. All the energy values mentioned in this article are per unit cell. 

\section{Topological Overprotection, Non-local String Operators and Emergent glassiness at finite temperature in Disorder free Kitaev model}\label{sec:Strings}
In this section, we elaborate on the discussion of the non-local string excitations in the Kitaev Honeycomb Lattice model and their constrained dynamics. The ground state of the uniform Kitaev model, a zero Fermion number sector, lies in the zero flux sector, as dictated by  Lieb's theorem (2D analogue of discussions in Sec. \ref{sec:Methods} of main text). This sector has full translational invariance, and a graphene-like Dirac cone spectrum of positive energy Fermion excitations (Majorana fermion) in Bloch states in the Brillouin zone of the honeycomb lattice. Other flux sectors bring in new physics. The spatial distribution of conserved and static $\pi$-fluxes, selected randomly from among the 2$^N$ sectors, is typically random. Consequently, one particle wave function of the positive energy Fermions will be non-Bloch-like and generically Anderson localized. 

We demonstrate that the Kitaev model has features that encourage glassiness at finite temperatures in the absence of disorder. The notion of Topological overprotection \cite{Chamon2005} induced glassiness was introduced by Chamon using 3D toric code quantum spin models. At the heart of Chamon's work is the observation that coupling of the constituent spin degree of freedom at lattice cites to a dissipative Bosonic (model thermal) bath results in the creation of defect (anyons) clusters. Because of topological protection, defect annihilation and propagation are severely constrained. It results in anomalous and slow relaxation - this is the beginning of glassiness.

To induce dynamics, following Chamon, we couple the constituent spin degrees of freedom at lattice sites to Bose oscillators (of some external heat bath) at every site: $a_i^{\alpha}, a_{i}^{\alpha \dagger}$ (this is analogous to applying site-dependent magnetic fields in Eq.~\ref{eq:ham}. And this discussion also applies in case of homogeneous fields mentioned in the main text).
\begin{equation*}
	H_{spin/bath} = \sum_{i,\alpha} g_\alpha S_i^\alpha \left(a_{i}^\alpha + a_{i}^{\alpha^\dagger}\right)
\end{equation*}

Where $g_\alpha$ is the coupling strength. It was shown \cite{Baskaran2007,Knolle2014} that a spin operator at site $i$, when acting on the ground state, creates a pair of static $\pi$-flux excitations in two plaquettes that share a single bond (in $\alpha$-direction) and a dynamical Majorana Fermion. During time evolution, the Majorana Fermion propagates away from the site $i$, while the two $\pi$-fluxes remain immobile. The dynamics of the two fluxes are restricted (topologically protected) in the following sense: They disappear only when a specific process takes place - when the nearest neighbour spin at a specific site creates/annihilates a (bath) boson and adds two more $\pi$-fluxes (thereby annihilating the two $\pi$-fluxes that are already present). If a different spin component at the same site $i$ creates/annihilates a boson, then the $\pi$-flux pair do not get annihilated but reoriented. If a wrong nearest neighbour spin creates/annihilates a bath Boson, two fluxes split and separate into two next nearest neighbour $\pi$-fluxes. 

Another extended operator arises from the liberated Majorana fermion. In terms of constituent spin operators, the Majorana fermion operator is a product of a string of spin operators. One end of the string is attached to the plaquette pair, and the other end carries the Majorana fermion. In other words, the Majorana fermion that has been created by coupling to bath Boson's degree of freedom is an extended object (strings). Strings of two Majorana Fermions can cross and get reconnected but never disappear. This feature of topological protection of strings is absent in models discussed in Ref. \onlinecite{Chamon2005}.

The above two types of non-local string operators from $\pi$-flux pairs and spin strings attached to Majorana fermions limit the disappearances of fluxes and discourage the proliferation of strings. Equilibration processes get slowed down, and glassiness may emerge. Thus, at any finite temperature, because of the production of $\pi$-flux excitations and strings, glassiness is induced via coupling to the bath.

From another point of view, the Quantum disentangled liquid \cite{Grover2014} character at any finite temperature is manifest and exact in the Kitaev spin liquid. We have thermally produced infinitely massive Z$_2$ fluxes, in the background of which light Majorana Fermions hop and attempt to delocalize. In the thermal ensemble, various superselected sectors with static fluxes appear and typically support Anderson localized positive energy neutral Fermions \cite{Giordano2021}. Thus we have overwhelming members of the thermal ensemble that form a quantum disentangled liquid with a high susceptibility for glassiness and non-thermalization.

\section{Exact diagonalization calculations}\label{sec:ED}

\begin{figure}[h] 
\centering
\includegraphics[width=0.8\linewidth]{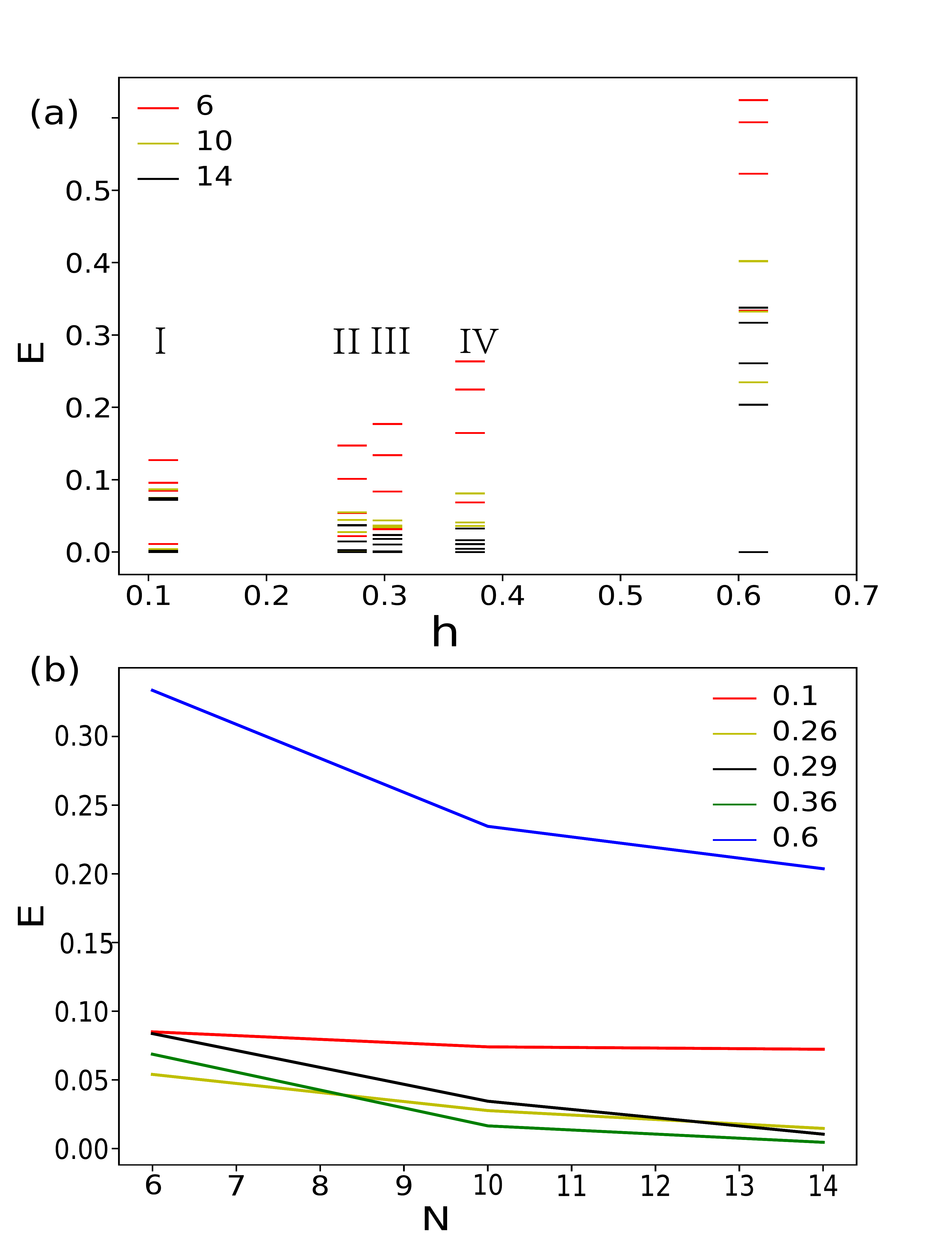}
\caption{(a) $E-E_0$ at the magnetic fields: h= 0.1 (Phase-I), 0.26 (Phase-II), 0.29 (Phase-III), 0.36 (Phase-IV) and 0.6 (Phase-V) for system sizes $N=$ 6, 10, and 14 are plotted. E$_0$ is the corresponding ground state energy. (b) The gap between the first excited state from the ground state at the above-mentioned magnetic fields is plotted as a function of system size ($N$).}
\label{fig:EDresults}
\end{figure}

We have performed the exact diagonalization (ED) calculations with the system sizes $N=$ 6, 10 and 14. The results are shown in Fig.~\ref{fig:EDresults}, where we have plotted the lowest five energy levels at five representative field values. A plot of the energy gap between the ground state and the first excited state is given in Fig.~\ref{fig:EDresults}(b). We find that in Phases II, III and IV, the energy gap decreases with increasing system size. The gap decreases much more slowly in the polarised phase (V), while that in the homogeneous phase (I) is nearly constant. There is a ground state degeneracy in Phases I and II, as also observed in the DMRG results. Within the numerically accessible system sizes, the ED results are consistent with our DMRG conclusions.


\bibliographystyle{apsrev4-2}
\bibliography{main.bib}

\end{document}